\documentclass{aastex631}

\usepackage{amsthm,amsmath,amssymb}
\usepackage{float}
\usepackage{url}
\usepackage{xcolor}

\begin{document}

\title{A Homogeneous Benchmark of Exoplanet Mass-Radius Relations}

\author[0000-0001-5768-0340]{Ailar Alizadehsabegh}
\affiliation{Department of Photonics, University of Tabriz, Tabriz, Iran}
\email{ailaralizadeh1999@gmail.com}

\author[0000-0001-9420-8790]{Soroush Sarabi}
\affiliation{Raderon AI Laboratory, BC., V5C 0J3 Burnaby, Canada}
\email{info@raderonlab.ca}

\author[0000-0002-0196-9732]{Atila Poro}
\affiliation{Astronomy Department of the Raderon AI Laboratory, BC., V5C 0J3 Burnaby, Canada}
\email{atilaporo@bsnp.info}

\begin{abstract}
The mass-radius ($M$-$R$) relation is fundamental to inferring the bulk composition and internal structure of exoplanets. However, published relations remain difficult to compare directly because studies adopt different planetary samples, fitting procedures, and evaluation metrics. We construct a homogeneous, quality-controlled benchmark of 963 exoplanets with dynamically measured masses and transit radii from the NASA Exoplanet Archive. We compare empirical power laws, probabilistic models, and machine-learning (ML) regressors using point accuracy, interval coverage, and probabilistic calibration. On the full test set, a multi-feature random forest has the lowest root-mean-square error, 0.329~dex in $\log_{10} M$ versus 0.368~dex for a fitted two-parameter power law ($p = 0.007$ unadjusted; $p = 0.059$ after Holm correction across 13 comparisons). Radius-only ML models perform comparably to the single power law. In three ranges of the measured radius, several methods improve on the power law below 1.6 $R_\oplus$, but none does between 1.6 and 4 $R_\oplus$. Including stellar and orbital covariates significantly improves predictions only for low-mass planets ($M < 4.4\,M_\oplus$), although this result is sensitive to missing covariate data and weakens when the training and test sets are separated by host star. Within their calibrated radius domains, the published empirical and probabilistic relations remain highly competitive. Finally, a first-order variance decomposition shows that formal measurement uncertainties account for only $\sim$3\% of the residual variance (0.067 of 0.401~dex), and the remaining $\sim$0.396~dex arises from compositional diversity, selection effects, and model approximation error.
\end{abstract}

\keywords{planetary systems, fundamental parameters, machine learning, methods: statistical}

\section{Introduction}
\label{sec:intro}
The planetary mass–radius relation provides a fundamental empirical framework for investigating the bulk composition and internal structure of exoplanets. Observational characterization, however, rarely yields both parameters from a single technique. Transit photometry constrains the planet-to-star radius ratio, from which the planetary radius is inferred using the stellar radius \citep{carter2011transit}[1]. Planetary masses, in turn, can be determined through radial velocity (RV) measurements, which yield $M_{\sin i}$ in the absence of an independent orbital inclination constraint, transit-timing variations (TTVs), which probe planetary masses through dynamical interactions, and gravitational microlensing, which primarily constrains the planet-to-host mass ratio \citep{2007ApJ...660..781B,cochran2011kepler,bouchy2011sophie}[2,3,4]. Even when both measurements are feasible for the same system, they seldom achieve comparable relative precision. Moreover, systematic offsets exist between mass measurement techniques, such as the lower bulk densities typically inferred from TTVs relative to RVs at longer orbital periods \citep{2016MNRAS.457.4384S,mills2017planetary}[5,6], and high precision RV follow-up remains concentrated around bright host stars \citep{batalha2010selection}[7]. These observational and computational limitations motivate the use of empirical $M$-$R$ relations to predict the unmeasured planetary parameter from the available observables \citep{1969ApJ...158..809Z,otegi2020revisited}[8,9].

Over the past decade, numerous $M$-$R$ relations have been proposed across three main methodological categories: deterministic parametric fits (primarily broken power laws), probabilistic models that parameterize the conditional mass distribution at a given radius, and machine-learning (ML) algorithms that fit non-parametric, multi-dimensional mappings. Differences in catalog versions, precision thresholds, fitting objectives, and evaluation metrics across studies make direct comparisons of their reported predictive accuracies difficult. In particular, it remains unclear how much predictive accuracy is gained by moving from low-dimensional physical parameterizations to flexible ML regressors across different planetary mass ranges, or how well these methods preserve uncertainty calibration.

In this study, we establish a homogeneous quantitative benchmark for exoplanet $M$-$R$ relations. First, we compile a quality-controlled sample of 963 planets with dynamically determined masses and transit-derived radii, applying explicit filters to avoid circularity associated with catalog-imputed values. Second, we evaluate representative empirical, probabilistic, and ML models on a common held-out test set using consistent point, interval, and probabilistic metrics stratified by planetary mass range. Third, we examine how discovery selection biases affect the $M$-$R$ residuals, evaluate distribution-free prediction intervals using conformal prediction, and assess the stability of model rankings under alternative data-quality cuts, mass boundaries, and feature sets. Section \ref{sec:review} reviews existing $M$-$R$ modeling approaches; Section \ref{sec:data} describes the construction of the benchmark dataset; Section \ref{sec:methods} details the evaluated models and statistical protocol; Section \ref{sec:results} presents the benchmark results; and Sections \ref{sec:discussion} and \ref{sec:conclusions} discuss the physical and methodological implications, limitations, and conclusions.

\vspace{0.5cm}
\section{A Critical Review of $M$-$R$ Approaches}
\label{sec:review}
Published $M$-$R$ relations differ primarily along four methodological dimensions: functional form, treatment of uncertainty, dimensionality of the input features, and the adoption of transition boundaries between planetary populations.

\subsection{Functional form}
\label{sec:review:form}
The most widely used parameterization is a piecewise power law of the form $R \propto M^{\beta}$ (or $M \propto R^{\alpha}$), in which the power-law exponent changes across distinct mass intervals \citep{weiss2013mass,weiss2014mass,hatzes2015definition,chen2016probabilistic,bashi2017two,otegi2020revisited,edmondson2023breaking,muller2024mass}[10,11,12,13,14,9,15,16]. Piecewise power laws are analytically tractable and reflect the expected physical scaling of compression and envelope accretion, but they impose a rigid functional shape with sharp slope discontinuities at the transition points. Non-parametric statistical models relax this assumption: the Bernstein-polynomial MRExo framework of \cite{kanodia2023beyond}[17], which builds on the formalism of \cite{ning2018predicting}[18], model the joint mass-radius distribution smoothly at the cost of requiring larger calibration samples. Supervised ML regressors \citep{2019A&A...630A.135U,tasker2020estimating,mousavi2023revisiting}[19,20,21] provide still greater functional flexibility in multi-dimensional spaces, though their fitted mappings are less directly interpretable in terms of interior structure. However, some previous ML studies have primarily investigated the prediction of planetary radii from mass and additional stellar or orbital parameters \citep{tasker2020estimating,mousavi2023revisiting}[20,21], whereas our benchmark, like the neural-network study of \cite{tasker2020estimating}[20], addresses the inverse problem of estimating planetary masses from radii.

\subsection{Treatment of uncertainty}
\label{sec:review:unc}
A second distinction separates deterministic point estimators from probabilistic models. Deterministic relations return a single predicted mass for a given radius and therefore neglect the substantial astrophysical dispersion observed at fixed radius. Probabilistic frameworks \citep{wolfgang2016probabilistic,chen2016probabilistic,parviainen2024spright}[22,13,23] model this conditional distribution explicitly, separating intrinsic scatter from observational measurement noise and, in hierarchical implementations, accounting for asymmetric error bars and upper limits \citep{kanodia2023beyond}[17]. For physical inference and survey planning, the calibration of the predicted uncertainty interval is as critical as the accuracy of the point estimate.

\subsection{Dimensionality and features}
\label{sec:review:dim}
Classical $M$-$R$ relations rely solely on mass and radius. Several recent studies incorporate additional stellar and orbital parameters to account for secondary physical processes, including equilibrium temperature to capture radius inflation in irradiated gas giants \citep{edmondson2023breaking}[15], stellar mass and insolation flux \citep{kanodia2023beyond}[17], and bulk density priors \citep{muller2024mass}[16]. While additional covariates can account for irradiation-driven radius variations, they also introduce missing-data patterns and survey selection dependencies that must be tested against a radius-only baseline.

\subsection{Mass ranges and transition boundaries}
\label{sec:review:ranges}
Most empirical and probabilistic studies divide the planetary population into three broad mass classes: rocky planets, volatile-rich intermediate planets (super-Earths and sub-Neptunes), and gas giants (Table~\ref{tab:transitions}). The transition from rocky to volatile-rich planets is typically placed between $\sim$4 and $10\,M_\oplus$, reflecting the threshold for retaining a significant $\mathrm{H/He}$ or volatile envelope, while the transition to gas giants occurs near $\sim$100 to $150\,M_\oplus$, where electron degeneracy pressure in the interior causes planetary radius to become nearly independent of mass. However, the exact transition masses and power-law slopes vary noticeably across studies due to differences in sample selection, measurement precision cuts, and regression methodology \citep{otegi2020revisited,edmondson2023breaking,muller2024mass}[9,15,16].

\begin{table*}
\centering
\caption{Representative reported mass classes, power-law slopes, and giant-planet transition boundaries from the literature.}
\small
\setlength{\tabcolsep}{3pt}
\begin{tabular}{lccc}
\hline
Study & Low-mass slope & Intermediate slope & Giant transition ($M_\oplus$) \\
\hline
\cite{weiss2014mass}[11] & $\sim$ density law & $M\propto R^{0.93}$ & - \\
\cite{bashi2017two}[14] & - & $R\propto M^{0.55}$ & $124\pm7$ \\
\cite{chen2016probabilistic}[13] & $R\propto M^{0.28}$ & $R\propto M^{0.59}$ & $\sim$132 \\
\cite{otegi2020revisited}[9] & $R\propto M^{0.29}$ & $R\propto M^{0.63}$ & $\sim$120 \\
\cite{muller2024mass}[16] & $R\propto M^{0.27}$ & $R\propto M^{0.67}$ & $127\pm7$ \\
\cite{edmondson2023breaking}[15] & $R\propto M^{0.28}$ & $R\propto M^{0.68}$ & $115\pm19$ \\
\hline
\end{tabular}
\label{tab:transitions}
\end{table*}

\subsection{Assumptions and limitations}
\label{sec:review:limits}
All $M$-$R$ relations face physical compositional degeneracy: a dense rocky super-Earth, an ocean world with a substantial water layer, and a rocky core surrounded by a tenuous $\mathrm{H/He}$ envelope can occupy identical locations in the $M$-$R$ plane. Consequently, no relation based solely on mass and radius can uniquely determine interior composition for intermediate-mass planets, which sets an intrinsic floor on predictive accuracy. Furthermore, all empirical calibrations inherit the detection biases of transit and RV surveys. Evaluating how different modeling families navigate these constraints requires testing them on an identical planetary sample under unified metrics, as detailed in Sections~\ref{sec:data} through \ref{sec:results}.

\vspace{0.5cm}
\section{The Benchmark Dataset}
\label{sec:data}
We compile our planetary sample from the NASA Exoplanet Archive Planetary Systems (PS) table \citep{akeson2013nasa,2025PSJ.....6..186C}[24,25], accessed via the Table Access Protocol (TAP) service on 2026 August 30. Because the archive is continuously updated as new planetary confirmations and parameter revisions are published, recording this exact retrieval date defines the fixed observational snapshot analyzed throughout this work.

\subsection{Sample selection and quality control}
\label{sec:data:qc}
The NASA Exoplanet Archive populates missing planetary masses or radii for many systems using built-in empirical $M$-$R$ conversions. Retaining such imputed entries in a benchmark would introduce circularity by evaluating $M$-$R$ models against values generated by another $M$-$R$ relation. We therefore restrict the sample to rows whose mass provenance flag indicates a direct mass determination, excluding minimum masses ($M\sin i$) and all catalog-calculated estimates. Furthermore, because directly imaged companions in the archive carry masses inferred from thermal evolution models at an assumed system age, we require a dynamical mass measurement derived from gravitational perturbations (via RV or TTV). This dynamical-mass requirement removes 30 directly imaged rows during initial filtering, corresponding to 11 objects that would otherwise pass our precision cuts; we evaluate the effect of retaining these 11 objects as a sensitivity test in Section~\ref{sec:results:robust}.

\begin{figure*}[ht]
\centering
\includegraphics[width=0.7\linewidth]{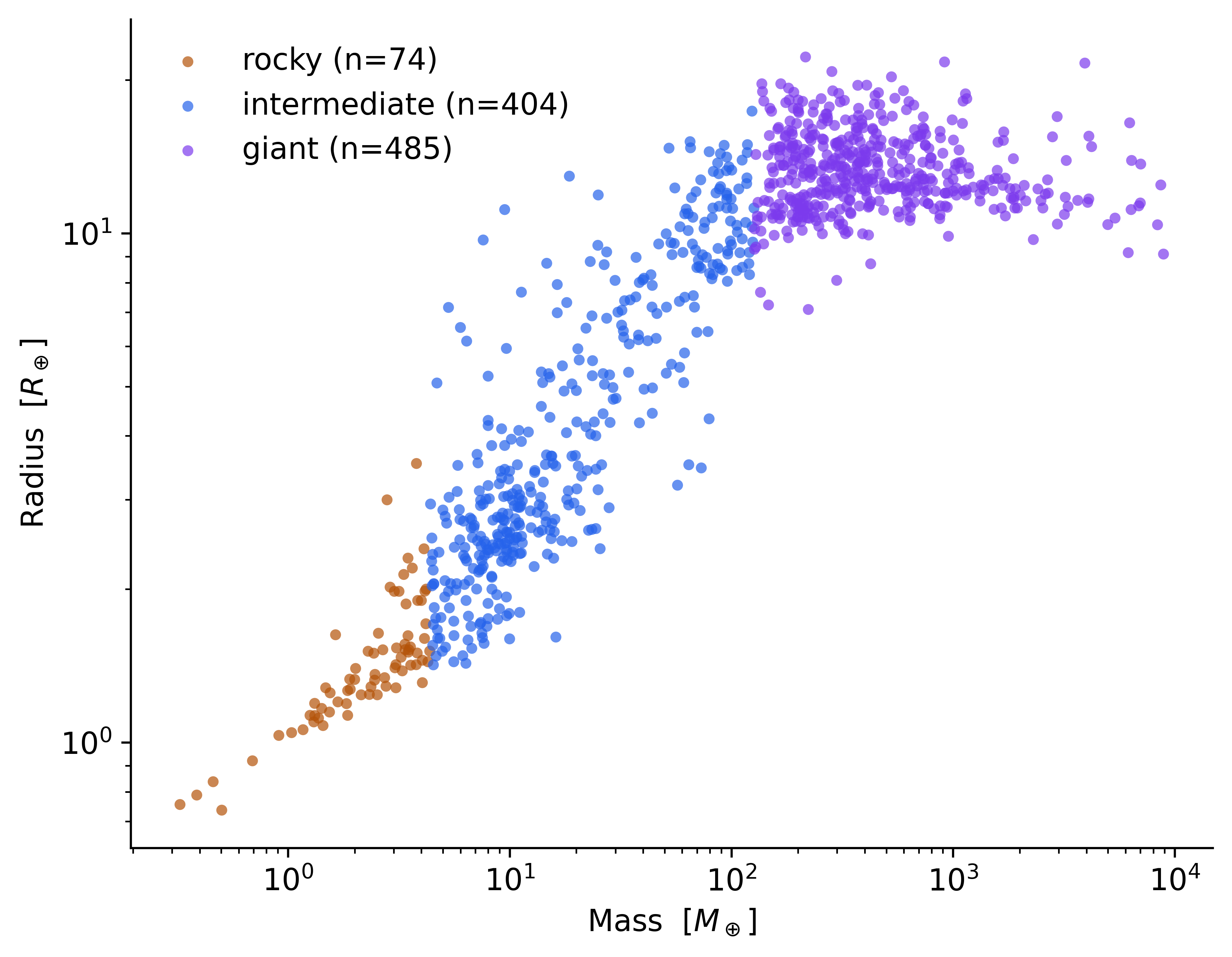}
\caption{Benchmark sample of 963 exoplanets in the $M$-$R$ plane, colored by mass class following the boundaries of \cite{muller2024mass}[16]. The intermediate mass range ($4.4 \le M < 127\,M_\oplus$), where interior compositional degeneracy is strong, exhibits a broad spread in mass at fixed radius. Because the benchmark predicts mass from radius, predictive difficulty is governed by the horizontal dispersion in $\log_{10} M$ at a given $\log_{10} R$.}
\label{fig:mrplane}
\end{figure*}

Starting from the non-controversial parameter solutions in the PS table, we require finite, positive mass and radius measurements within $0 < R < 30\,R_\oplus$ and impose fiducial relative-uncertainty thresholds of $\sigma_M/M < 0.25$ and $\sigma_R/R < 0.08$, where $\sigma_M$ and $\sigma_R$ denote the mean of the asymmetric upper and lower error bars. When multiple parameter solutions survive for a single planet, we select the entry that minimizes the combined relative uncertainty $\sigma_M/M + \sigma_R/R$. To account for asymmetric error bars in log space, we draw $10^{3}$ Monte Carlo realizations of $M$ and $R$ per planet from two-sided Gaussian distributions matched to the catalog upper and lower uncertainties, recording the resulting mean and standard deviation in $\log_{10}$ units. These fiducial quality cuts yield a final benchmark sample of 963 exoplanets, which we partition into 723 training planets (75\%) and 240 held-out test planets (25\%) using a fixed random seed (42) stratified by mass class. Because the split is made at the planet level, planets of the same system can fall in different sets. In the fiducial split, 27 host stars have planets in both sets, and 33 of the 240 test planets share their host star with at least one training planet. We test the effect of this overlap in Section \ref{sec:results:robust} with splits grouped by host star. Figure \ref{fig:mrplane} displays the distribution of the benchmark sample in the $M$-$R$ plane.

\subsection{Mass classes and boundaries}
\label{sec:data:classes}
To evaluate performance across distinct regions of parameter space, we adopt the literature mass boundaries of \cite{muller2024mass}[16]: low-mass ($M < 4.4\,M_\oplus$, 74 planets), intermediate-mass ($4.4 \le M < 127\,M_\oplus$, 404 planets), and giant ($M \ge 127\,M_\oplus$, 485 planets). These intervals represent empirical mass cuts and do not correspond one-to-one with bulk composition. While the 74 planets in the $M < 4.4\,M_\oplus$ bin have a median bulk density of $5.0\text{ g cm}^{-3}$ typical of terrestrial bodies, seven have bulk densities below $2.0\text{ g cm}^{-3}$ and 14 have radii exceeding $1.8\,R_\oplus$, indicating a small admixture of low-mass volatile-rich planets. We retain the published mass thresholds of \cite{muller2024mass}[16] for consistency with prior work and test alternative boundary definitions in Section~\ref{sec:results:robust}. Because these three classes are defined by catalog mass, they serve as diagnostic categories for evaluating test-set errors; in forward applications where mass is unknown, radius-based intervals (such as $R < 1.6\,R_\oplus$, $1.6 \le R \le 4.0\,R_\oplus$, and $R > 4.0\,R_\oplus$) must be used instead, and Section \ref{sec:results:ranges} reports the benchmark in these radius ranges.

\vspace{0.5cm}
\section{Methods and Evaluation Framework}
\label{sec:methods}

\subsection{Evaluated models}
\label{sec:methods:methods}
We evaluate four groups of models, all configured to predict planetary mass from radius ($R \to M$). The first group comprises empirical parametric relations from the literature \citep{weiss2014mass,bashi2017two,otegi2020revisited}[11,14,9], each evaluated within its adopted validity domain (with exact equations, unit conversions, and domain limits detailed in Appendix~\ref{app:empirical}). The second group consists of three simple reference baselines fitted to the training split: a sample-mean predictor (Baseline-Mean), a constant bulk-density scaling (Baseline-ConstDensity), and an ordinary-least-squares single power law in log space (Baseline-PowerLaw), along with its Gaussian constant-variance probabilistic counterpart (Prob-PowerLaw). The third group includes published probabilistic packages evaluated directly through their public interfaces: Forecaster \citep{chen2016probabilistic}[13] and SPRIGHT \citep{parviainen2024spright}[23], while MRExo is discussed in Appendix~\ref{app:empirical}. The fourth group comprises four supervised ML regressors implemented in scikit-learn \citep{pedregosa2011scikit}[26]: random forest (RandomForest), gradient boosting (GradientBoosting), support-vector regression (SVR), and a multilayer perceptron neural network (MLP), all trained on $\log_{10} R$ alone. In addition, we train a multi-feature random forest (RandomForest-multi) using planetary radius, equilibrium temperature, insolation flux, stellar mass, and stellar effective temperature to test the predictive contribution of host-star and orbital covariates.

\subsection{Unified evaluation protocol}
\label{sec:methods:protocol}
All models are scored in $\log_{10}(M/M_\oplus)$ on the same held-out test set of 240 planets. For the multi-feature model, missing values in the secondary covariates (equilibrium temperature, insolation, stellar mass, and stellar effective temperature) are imputed using the training-set median within the model pipeline so that all 240 test planets are retained. Each ML regressor is tuned via five-fold cross-validation restricted strictly to the 723-planet training set, refitted on the full training set using the selected hyperparameters, and evaluated once on the held-out test set. The hyperparameter search grids and selected configurations are listed in Appendix~\ref{app:hyper}. Pre-trained probabilistic packages \citep{chen2016probabilistic,parviainen2024spright}[13,23] are evaluated without refitting.

To equip the point ML models with distribution-free uncertainty intervals, we apply split conformal prediction \citep{angelopoulos2023conformal}[27]. The 723-planet training set is partitioned into a 70\% proper training subset and a 30\% calibration subset; absolute residuals on the calibration subset define the empirical quantiles used to construct 68\% and 95\% prediction intervals on the test set. Under the assumption of exchangeability between calibration and test samples, split conformal prediction guarantees finite-sample marginal coverage across the full test distribution, though it does not guarantee conditional coverage within individual mass ranges. We also use permutation importance on the held-out test set to quantify feature attribution in RandomForest-multi, measuring the increase in test RMSE when each input feature is randomly permuted (an approach that is model-agnostic, unlike tree-specific or local SHapley Additive exPlanations (SHAP) approximations; \citep{lundberg2017unified}[28]).

\subsection{Performance metrics and statistical testing}
\label{sec:methods:metrics}
We evaluate point predictions using root-mean-square error (RMSE) and the coefficient of determination ($R^2$) in $\log_{10} M$. Uncertainty intervals are assessed via prediction-interval coverage probability (PICP, the fraction of test planets whose true masses fall within the nominal 68\% and 95\% intervals) and mean interval width in dex. Full predictive distributions are scored using the continuous ranked probability score (CRPS, in dex of $\log_{10} M$), negative log predictive density (NLPD), and the probability integral transform (PIT) calibration error. The PIT error is defined as the mean absolute deviation of a ten-bin PIT histogram from a uniform distribution, normalized by the expected count per bin; values near zero indicate well-calibrated posteriors, whereas values near unity indicate severe miscalibration. Because histogram fluctuations depend on sample size, PIT values must be compared on identical planetary subsets.

Because all competing models are evaluated on the same test planets, pairwise model comparisons are conducted using paired resampling. Confidence intervals (95\%) on RMSE and $\Delta\mathrm{RMSE} = \mathrm{RMSE}(\mathrm{method}) - \mathrm{RMSE}(\mathrm{baseline})$ are computed from $2{,}000$ paired percentile bootstrap resamples of the test set. Statistical significance for pairwise differences is evaluated using a two-sided paired sign-flip permutation test with $10{,}000$ permutations of the per-planet squared errors. To account for multiple hypothesis testing across the benchmark, we apply the Holm step-down correction across two pre-defined comparison families: the 13 method-versus-baseline comparisons in Table~\ref{tab:lookup}, and the 4 comparisons (full test set plus three mass classes) testing the covariate gain of RandomForest-multi over the radius-only RandomForest. The same two families are used in the three radius ranges of Section \ref{sec:results:ranges}, with the methods that can be scored in each range and with the three radius ranges in place of the mass classes. When evaluating domain-restricted models, both the model and its baseline comparator are scored on the exact intersection of test planets within the valid radius domain.

\vspace{0.5cm}
\section{Results}
\label{sec:results}

\subsection{Overall model performance}
\label{sec:results:overall}
Table~\ref{tab:lookup} presents the test-set performance for all evaluated methods, separated into full-coverage models ($N = 240$ test planets) and domain-restricted literature relations. Among full-coverage methods, RandomForest-multi achieves the lowest overall RMSE at $0.329\text{ dex}$ ($95\%$ bootstrap CI $[0.282,\,0.380]$, $R^2 = 0.859$), compared with $0.368\text{ dex}$ ($[0.32,\,0.42]$, $R^2 = 0.82$) for the fitted two-parameter power law (Baseline-PowerLaw). In a paired comparison across all 240 test planets, RandomForest-multi reduces RMSE relative to Baseline-PowerLaw by $\Delta\mathrm{RMSE} = -0.038\text{ dex}$ ($95\%$ CI $[-0.068,\,-0.011]$). The paired sign-flip permutation test yields an unadjusted $p = 0.007$, which increases to $p = 0.059$ after Holm correction across the 13 method-versus-baseline comparisons in Table~\ref{tab:lookup}. Thus, while the point improvement of the multi-feature random forest over a simple power law is consistent in sign, its magnitude on a test sample of 240 planets lies just above the conventional $\alpha = 0.05$ significance threshold once multiple testing is accounted for.

When ML regressors are restricted to planetary radius alone (GradientBoosting, RandomForest, MLP, and SVR), their overall RMSE values span $0.344$ to $0.353\text{ dex}$, and none differs significantly from Baseline-PowerLaw even before multiplicity correction (unadjusted $p \ge 0.067$). This result is corroborated by the radius-only row of Table~\ref{tab:robust}, where the leading radius-only ML model differs from the power law by $\Delta\mathrm{RMSE} = -0.020\text{ dex}$ ($95\%$ CI $[-0.043,\,0.003]$). Flexible non-parametric regression on radius alone therefore provides no statistically distinguishable gain over a single log-linear fit.

To test whether the lower error of RandomForest-multi stems directly from the added stellar and orbital covariates, we perform a paired comparison against the radius-only RandomForest algorithm. Over the full test set, the paired difference is $\Delta\mathrm{RMSE} = -0.015\text{ dex}$ ($95\%$ CI $[-0.030,\,0.003]$, Holm-adjusted $p = 0.260$). Stratified by mass class, the improvement is concentrated entirely in the low-mass range ($M < 4.4\,M_\oplus$), where $\Delta\mathrm{RMSE} = -0.028\text{ dex}$ ($95\%$ CI $[-0.042,\,-0.010]$, raw $p = 0.011$, Holm-adjusted $p = 0.043$), whereas the intervals for intermediate-mass and giant planets both include zero. Even within the low-mass group, however, the physical interpretation of this covariate gain is limited by catalog completeness: stellar insolation is missing for 64\% of the overall test set and 44\% of the low-mass test planets, requiring median imputation. Repeating the paired test on the 80 test planets that have complete measurements for all covariates yields $\Delta\mathrm{RMSE} = -0.006\text{ dex}$ ($[-0.037,\,0.035]$, $p = 0.78$), and only eight of the 18 low-mass test planets possess complete covariates. Consequently, while a small covariate improvement among low-mass planets is physically plausible given irradiation-driven atmospheric mass loss, a larger sample with uniform stellar and orbital characterization is required to separate true physical dependence from catalog missingness patterns. The low-mass gain also weakens when the training and test sets are separated by host star (Section 5.4), since planets of the same system share their stellar parameters.

In the lower block of Table~\ref{tab:lookup}, the domain-restricted empirical relations are scored only on the fraction of test planets falling within their valid radius ranges (indicated by the coverage column). The relation of \cite{weiss2014mass}[11], valid for $0.5 \le R \le 4.0\,R_\oplus$, covers 33\% of the test set and achieves an RMSE of $0.197\text{ dex}$ on that subset, while SPRIGHT covers 40\% ($0.5 \le R \le 6.5\,R_\oplus$) with an RMSE of $0.207\text{ dex}$. Because these restricted subsets exclude massive gas giants where mass scatter at fixed radius is largest, their summary RMSE values cannot be compared directly with full-coverage scores without matching domains.

\begin{table*}
\centering
\caption{Mass-prediction performance on the held-out test set in dex of $\log_{10} M$, with 95\% percentile bootstrap confidence intervals in square brackets. The coverage column (Cov.) lists the fraction of the 240 test planets within each model's valid radius domain. Full-coverage and domain-restricted blocks are evaluated on different planetary subsets and are not directly comparable without domain matching. Empirical relations follow \cite{weiss2014mass}[11], \cite{bashi2017two}[14], and \cite{otegi2020revisited}[9] as detailed in Appendix~\ref{app:empirical}.}
\footnotesize
\setlength{\tabcolsep}{2.5pt}
\begin{tabular}{llcccccc}
\hline
Method & Family & Cov. & RMSE & $R^2$ & RMSE$_{\rm low}$ & RMSE$_{\rm int}$ & RMSE$_{\rm giant}$ \\
\hline
\multicolumn{8}{c}{Scored on the full test set ($N = 240$)} \\
\hline
RandomForest-multi & ML & 1.00 & 0.329\,[0.28,0.38] & 0.86 & 0.191\,[0.13,0.25] & 0.299\,[0.24,0.36] & 0.367\,[0.29,0.44] \\
GradientBoosting & ML & 1.00 & 0.344\,[0.29,0.40] & 0.85 & 0.209\,[0.14,0.28] & 0.310\,[0.25,0.38] & 0.384\,[0.31,0.46] \\
RandomForest & ML & 1.00 & 0.344\,[0.29,0.40] & 0.85 & 0.219\,[0.15,0.29] & 0.315\,[0.25,0.38] & 0.380\,[0.30,0.46] \\
MLP & ML & 1.00 & 0.349\,[0.30,0.40] & 0.84 & 0.207\,[0.13,0.27] & 0.331\,[0.26,0.40] & 0.379\,[0.30,0.45] \\
SVR & ML & 1.00 & 0.353\,[0.30,0.41] & 0.84 & 0.207\,[0.14,0.27] & 0.299\,[0.24,0.37] & 0.408\,[0.33,0.48] \\
Baseline-PowerLaw & baseline & 1.00 & 0.368\,[0.32,0.42] & 0.82 & 0.158\,[0.11,0.20] & 0.339\,[0.28,0.40] & 0.410\,[0.33,0.49] \\
Prob-PowerLaw & probabilistic & 1.00 & 0.368\,[0.32,0.42] & 0.82 & 0.158\,[0.11,0.20] & 0.339\,[0.28,0.40] & 0.410\,[0.33,0.49] \\
Baseline-ConstDensity & baseline & 1.00 & 0.481\,[0.44,0.52] & 0.70 & 0.633\,[0.55,0.71] & 0.512\,[0.45,0.57] & 0.426\,[0.36,0.49] \\
Forecaster \citep{chen2016probabilistic}[13] & probabilistic & 1.00 & 0.657\,[0.59,0.72] & 0.44 & 0.142\,[0.09,0.19] & 0.614\,[0.47,0.74] & 0.734\,[0.67,0.80] \\
Baseline-Mean & baseline & 1.00 & 0.878\,[0.83,0.93] & -0.00 & 1.466\,[1.39,1.55] & 0.814\,[0.76,0.87] & 0.810\,[0.74,0.88] \\
\hline
\multicolumn{8}{c}{Domain-restricted: scored within published validity domain} \\
\hline
\cite{weiss2014mass}[11] & empirical & 0.33 & 0.197\,[0.17,0.23] & 0.48 & 0.142\,[0.08,0.20] & 0.210\,[0.18,0.25] & - \\
SPRIGHT & probabilistic & 0.40 & 0.207\,[0.16,0.25] & 0.66 & 0.149\,[0.09,0.21] & 0.218\,[0.17,0.27] & - \\
\cite{bashi2017two}[14] (inverted) & empirical & 0.64 & 0.435\,[0.35,0.52] & 0.72 & 0.144\,[0.11,0.18] & 0.237\,[0.20,0.28] & 0.774\,[0.61,0.94] \\
\cite{otegi2020revisited}[9] (radius-only) & empirical & 0.85 & 0.538\,[0.47,0.61] & 0.63 & 0.147\,[0.09,0.19] & 0.222\,[0.18,0.26] & 0.792\,[0.70,0.89] \\
\hline
\end{tabular}
\label{tab:lookup}
\end{table*}

\begin{table*}
\centering
\caption{Mass-prediction RMSE in three ranges of the measured radius on the held-out test set, in dex of $\log_{10}M$, with 95\% percentile bootstrap intervals in square brackets. The ranges contain 14, 65 and 161 test planets. Domain-restricted relations are scored only on the planets they can predict, and where this is not the whole range the covered fraction is given in parentheses.}
\small
\begin{tabular}{lccc}
\hline\hline
Method & $R<1.6\,R_\oplus$ & $1.6\le R\le4\,R_\oplus$ & $R>4\,R_\oplus$ \\
\hline
\multicolumn{4}{c}{Scored on the full test set} \\
\hline
RandomForest-multi & 0.121\,[0.08,0.16] & 0.174\,[0.14,0.20] & 0.385\,[0.32,0.45] \\
GradientBoosting & 0.121\,[0.08,0.15] & 0.186\,[0.15,0.22] & 0.401\,[0.34,0.46] \\
RandomForest & 0.157\,[0.11,0.20] & 0.191\,[0.16,0.22] & 0.399\,[0.33,0.46] \\
MLP & 0.105\,[0.07,0.14] & 0.195\,[0.16,0.23] & 0.406\,[0.34,0.47] \\
SVR & 0.125\,[0.08,0.16] & 0.183\,[0.15,0.21] & 0.414\,[0.35,0.48] \\
Baseline-PowerLaw & 0.181\,[0.12,0.23] & 0.195\,[0.15,0.24] & 0.428\,[0.37,0.49] \\
Prob-PowerLaw & 0.181\,[0.12,0.23] & 0.195\,[0.15,0.24] & 0.428\,[0.37,0.49] \\
Baseline-ConstDensity & 0.789\,[0.73,0.84] & 0.467\,[0.41,0.53] & 0.451\,[0.39,0.50] \\
Forecaster \citep{chen2016probabilistic}[13] & 0.157\,[0.10,0.20] & 0.189\,[0.16,0.22] & 0.792\,[0.72,0.87] \\
Baseline-Mean & 1.436\,[1.33,1.55] & 1.027\,[0.97,1.08] & 0.737\,[0.67,0.80] \\
\hline
\multicolumn{4}{c}{Domain-restricted: scored within the published validity domain} \\
\hline
\cite{weiss2014mass}[11] & 0.087\,[0.06,0.11] & 0.213\,[0.18,0.25] & \nodata \\
SPRIGHT & 0.141\,[0.10,0.18] & 0.187\,[0.15,0.22] & 0.302\,[0.14,0.44] (0.11) \\
\cite{bashi2017two}[14] (inverted) & 0.192\,[0.13,0.24] & 0.196\,[0.16,0.24] & 0.591\,[0.46,0.71] (0.47) \\
\cite{otegi2020revisited}[9] (radius-only) & 0.147\,[0.09,0.19] & 0.184\,[0.15,0.22] & 0.672\,[0.59,0.76] (0.78) \\
\hline
\end{tabular}
\label{tab:radius}
\end{table*}

\subsection{Performance across mass and radius ranges}
\label{sec:results:ranges}
\begin{figure*}[ht]
\centering
\includegraphics[width=0.95\linewidth]{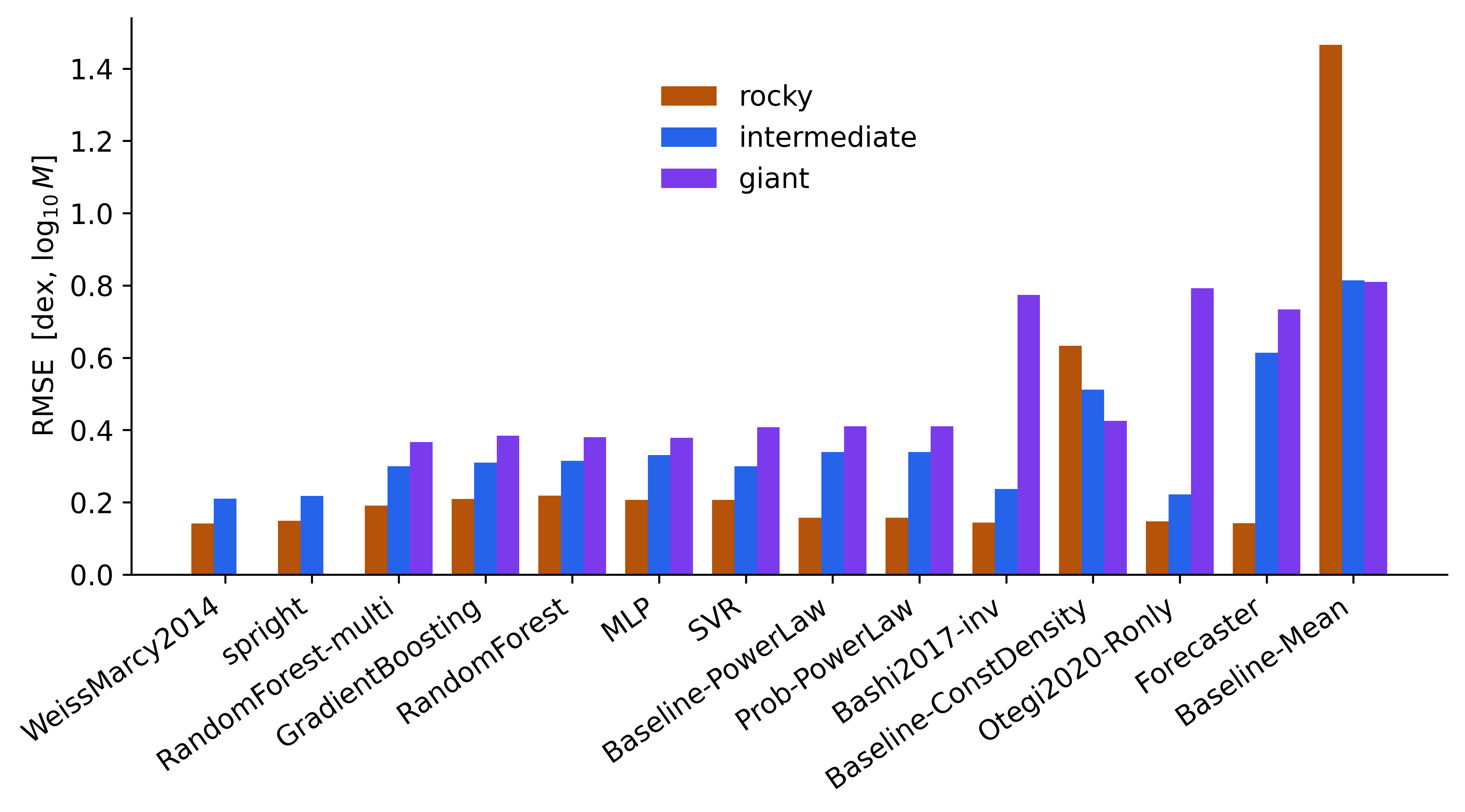}
\caption{Mass-prediction RMSE stratified by method and planetary mass class. Because relative model rankings shift across the low-mass, intermediate-mass, and giant intervals, a single global RMSE does not fully characterize model performance.}
\label{fig:rmse}
\end{figure*}

Figure~\ref{fig:rmse} and the right-hand columns of Table~\ref{tab:lookup} show how prediction errors vary across the three planetary mass classes. In the low-mass range ($M < 4.4\,M_\oplus$, $N = 18$ test planets), RMSE point estimates for the empirical, probabilistic, and ML models span $0.142$ to $0.219\text{ dex}$ with wide, mutually overlapping bootstrap confidence intervals, precluding a statistically separable winner. Across the full test set, eight of the ten full-coverage methods incur their largest RMSE in the giant-planet range ($M \ge 127\,M_\oplus$), exceeding their errors in the intermediate range ($4.4 \le M < 127\,M_\oplus$). Although compositional degeneracy between rocky cores, water worlds, and volatile envelopes is most pronounced at intermediate masses, that class spans a bounded interval in $\log_{10} M$ that limits maximum prediction errors. Above $\sim 127\,M_\oplus$, by contrast, degenerate interior compression causes planetary radius to plateau \citep{1969ApJ...158..809Z,bashi2017two}[8,14], so objects spanning more than an order of magnitude in mass occupy nearly identical radii; predicting mass from radius in the giant range is therefore inherently poorly conditioned. Only the uninformative Baseline-Mean and Baseline-ConstDensity models exhibit their largest errors in the low-mass range.

Stratifying by mass range also clarifies the behavior of the mass-bounded empirical relations of \cite{bashi2017two}[14] and \cite{otegi2020revisited}[9]. The radius-only adaptation of \cite{otegi2020revisited}[9], originally calibrated on planets below $120\,M_\oplus$, achieves an RMSE of $0.147\text{ dex}$ in the low-mass class and $0.222\text{ dex}$ in the intermediate class, outperforming all ML models in both intervals, but degrades to $0.792\text{ dex}$ on giant planets. The inverted relation of \cite{bashi2017two}[14] shows a nearly identical pattern ($0.144$, $0.237$, and $0.774\text{ dex}$ across the low, intermediate, and giant classes). Their elevated aggregate RMSE values ($0.538$ and $0.435\text{ dex}$) arise because a calibration upper limit defined in mass ($120$ or $124\,M_\oplus$) cannot exclude all massive gas giants when predictions are conditioned on radius alone, as inflated sub-Saturns and massive Jovian planets overlap substantially in radius between $8$ and $14\,R_\oplus$.

Because the mass classes are defined by the quantity being predicted, we also scored all methods in three ranges of the measured radius, $R<1.6\,R_\oplus$, $1.6\le R\le4\,R_\oplus$ and $R>4\,R_\oplus$, which contain 14, 65 and 161 test planets (Table \ref{tab:radius}). These ranges overlap only partly with the mass classes: 3 of the 14 planets below $1.6\,R_\oplus$ are intermediate-mass planets, 7 of the 65 planets between 1.6 and $4\,R_\oplus$ are low-mass planets, and 40 of the 161 planets above $4\,R_\oplus$ are intermediate-mass planets. Below $1.6\,R_\oplus$, the relation of \cite{weiss2014mass}[11] has the lowest error (0.087~dex). This relation, MLP, SVR, SPRIGHT and the relation of \cite{otegi2020revisited}[9] are significantly better than the fitted power law (0.181~dex) after a Holm correction within the range. Between 1.6 and $4\,R_\oplus$, where most sub-Neptunes lie, the ML models and the published relations have errors of 0.17 to 0.21~dex, and none differs significantly from the power law (0.195~dex). Above $4\,R_\oplus$ the errors are about twice as large (0.38 to 0.43~dex for the ML models and the power law), no method is significantly better than the power law after correction, and the mass-limited relations of \cite{bashi2017two}[14] and \cite{otegi2020revisited}[9], as well as Forecaster, have errors of 0.59 to 0.79~dex. RandomForest-multi improves on the radius-only RandomForest below $4\,R_\oplus$, by 0.036~dex below $1.6\,R_\oplus$ and 0.017~dex between 1.6 and $4\,R_\oplus$, with paired intervals that exclude zero. As for the low-mass class, this gain is less consistent when planets are grouped by host star, with a lower RMSE in 93 and 74 of the 100 grouped splits of Section 5.4. In practice, the choice of method therefore matters most for planets smaller than $1.6\,R_\oplus$ and least between 1.6 and $4\,R_\oplus$.

\subsection{Probabilistic calibration and interval coverage}
\label{sec:results:calib}
\begin{figure}[ht]
\centering
\includegraphics[width=0.5\linewidth]{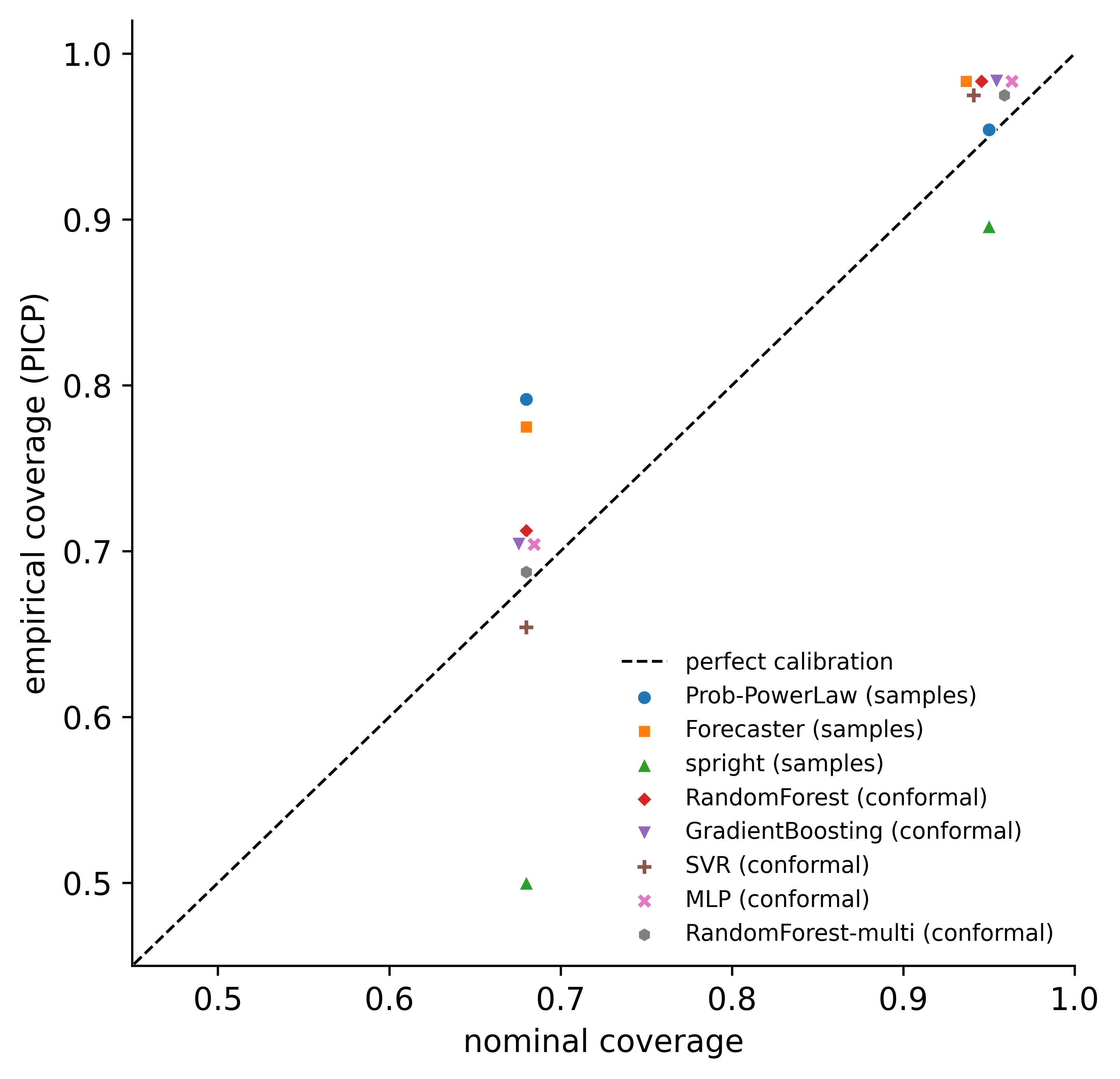}
\caption{Empirical versus nominal interval coverage on the test set, with horizontal offsets applied for visual clarity. Conformal ML intervals lie close to the diagonal across the full test set ($N = 240$), whereas SPRIGHT is evaluated on the 96-planet subset within its validity domain ($0.5 \le R \le 6.5\,R_\oplus$).}
\label{fig:calib}
\end{figure}

Figure~\ref{fig:calib} and Table~\ref{tab:prob} summarize the probabilistic and interval calibration results. Averaged across the five ML regressors equipped with split conformal prediction, empirical coverage on the test set is $0.69$ (ranging from $0.65$ to $0.71$ across models) at the nominal 68\% level and $0.98$ ($0.97$ to $0.98$) at the nominal 95\% level. While marginal coverage closely matches the nominal target as guaranteed by split conformal calibration, coverage varies systematically across mass classes: at the nominal 68\% level, all five conformal ML models over-cover low-mass planets ($0.78$ to $0.94$, with individual Wilson binomial intervals spanning $0.55$ to $0.99$ for $N = 18$) and under-cover giant planets ($0.55$ to $0.66$, Wilson intervals $0.46$ to $0.74$ for $N = 121$). This drift occurs because standard split conformal calibration applies a single marginal residual quantile across mass classes whose intrinsic residual dispersions differ. A similar pattern is observed when the test set is stratified by radius, which is known before a prediction is made: the 68\% conformal intervals over-cover below $4,R_\oplus$ (0.88 to 1.00) and under-cover above $4,R_\oplus$ (0.53 to 0.61). Achieving uniform conditional coverage in forward applications requires stratifying the calibration sample by observables known prior to prediction, such as radius bins, via class-conditional (Mondrian) conformal prediction \citep{vovk2005algorithmic,angelopoulos2023conformal}[29,27].

Among the models that output full predictive distributions, Prob-PowerLaw attains the lowest CRPS ($0.200\text{ dex}$) and PIT calibration error ($0.34$) over the full 240-planet test set, compared with a CRPS of $0.319\text{ dex}$ and PIT error of $0.38$ for Forecaster. However, SPRIGHT is defined only for $0.5 \le R \le 6.5\,R_\oplus$, returning posteriors for 96 test planets (40\% of the test set) that exclude the high-scatter giant population. When all three probabilistic models are evaluated on this common subset of 96 small and intermediate planets, SPRIGHT and Forecaster achieve nearly identical CRPS values of $0.113$ and $0.117\text{ dex}$ (paired difference $0.004\text{ dex}$, $95\%$ CI $[-0.005,\,0.014]$), and both significantly outperform Prob-PowerLaw ($0.141\text{ dex}$, with paired differences of $-0.028\text{ dex}$, $[-0.036,\,-0.019]$ for SPRIGHT and $-0.023\text{ dex}$, $[-0.036,\,-0.012]$ for Forecaster).

On the same 96-planet common domain, the three probabilistic models exhibit distinct interval calibration behavior. SPRIGHT produces narrow credible intervals (mean 68\% width of $0.27\text{ dex}$) that under-cover the catalog masses ($\mathrm{PICP}_{68} = 0.50$, $\mathrm{PICP}_{95} = 0.90$), whereas Forecaster yields slightly conservative intervals ($\mathrm{PICP}_{68} = 0.77$, $\mathrm{PICP}_{95} = 0.98$, with mean 68\% width of $0.46\text{ dex}$). By contrast, Prob-PowerLaw applies a global residual variance fitted across all masses and is therefore overly wide on small planets ($\mathrm{PICP}_{68} = 0.95$, width $0.82\text{ dex}$), yielding a worse common-domain PIT error ($0.71$) than either Forecaster ($0.50$) or SPRIGHT ($0.53$). The larger full-sample CRPS and interval width of Forecaster in Table~\ref{tab:prob} thus stem from its broad posterior across the degenerate giant-planet branch.

\begin{table*}
\centering
\caption{Probabilistic and interval evaluation metrics on the held-out test set. CRPS and interval widths carry units of dex in $\log_{10} M$, NLPD is the negative log predictive density, and PIT err.\ is the ten-bin probability integral transform calibration error defined in Section~\ref{sec:methods:metrics}. PICP gives the empirical fraction of test masses falling within the nominal 68\% and 95\% intervals.}
\small
\setlength{\tabcolsep}{3pt}
\begin{tabular}{lcccccccc}
\hline
Method & Cov. & CRPS & NLPD & PIT err. & PICP$_{68}$ & Width$_{68}$ & PICP$_{95}$ & Width$_{95}$ \\
\hline
\multicolumn{9}{c}{Models returning full predictive distributions} \\
\hline
SPRIGHT & 0.40 & 0.113 & -0.230 & 0.53 & 0.50 & 0.27 & 0.90 & 0.62 \\
Prob-PowerLaw & 1.00 & 0.200 & 0.430 & 0.34 & 0.79 & 0.82 & 0.95 & 1.61 \\
Forecaster \citep{chen2016probabilistic}[13] & 1.00 & 0.319 & 0.632 & 0.38 & 0.78 & 1.33 & 0.98 & 2.03 \\
\hline
\multicolumn{9}{c}{Point ML models with split-conformal intervals} \\
\hline
RandomForest & 1.00 & \nodata & \nodata & \nodata & 0.71 & 0.59 & 0.98 & 1.76 \\
GradientBoosting & 1.00 & \nodata & \nodata & \nodata & 0.70 & 0.60 & 0.98 & 1.85 \\
SVR & 1.00 & \nodata & \nodata & \nodata & 0.65 & 0.56 & 0.97 & 1.68 \\
MLP & 1.00 & \nodata & \nodata & \nodata & 0.70 & 0.62 & 0.98 & 1.77 \\
RandomForest-multi & 1.00 & \nodata & \nodata & \nodata & 0.69 & 0.54 & 0.97 & 1.61 \\
\hline
\end{tabular}
\label{tab:prob}
\end{table*}

\subsection{Sensitivity and split stability}
\label{sec:results:robust}
Table~\ref{tab:robust} reports the sensitivity of the benchmark across nine pipeline variations: the fiducial sample evaluated with fixed hyperparameters, stricter ($\sigma_M/M < 0.20$, $\sigma_R/R < 0.06$) and looser ($0.35 / 0.12$) uncertainty cuts, two alternative mass-boundary definitions (the transition boundaries of \citep{chen2016probabilistic}[13] and a wider low-mass threshold), retention of the 11 directly imaged planets, two restricted feature sets (radius only, and stellar parameters without irradiation covariates), and a training/test split grouped by host star. Across these perturbations, sample size ranges from $N = 759$ to $1{,}161$ planets; RandomForest-multi ranks first in seven configurations and MLP in two. In all nine cases, the best full-coverage ML model achieves an RMSE between $0.020$ and $0.076\text{ dex}$ below the fitted power law, with the 95\% paired bootstrap interval overlapping zero in three variants. Retaining the 11 directly imaged planets ($N = 974$) degrades the power-law RMSE from $0.368$ to $0.414\text{ dex}$, reflecting the systematic offset of evolutionary-model mass estimates at large radii (Section \ref{sec:data}).

To verify that our conclusions do not depend on the single random seed used for the fiducial 75/25 train-test partition, we repeated the stratified split 100 times, refitting Baseline-PowerLaw, RandomForest, and RandomForest-multi on each partition. Across the 100 partitions, RandomForest-multi yielded a lower RMSE than Baseline-PowerLaw in all 100 splits (median $\Delta\mathrm{RMSE} = -0.048\text{ dex}$, $2.5$ to $97.5$ percentile range $[-0.073,\,-0.027]$), the radius-only RandomForest had a lower RMSE than Baseline-PowerLaw in all 100 splits (median $-0.024\text{ dex}$, $[-0.047,\,-0.008]$), and RandomForest-multi outperformed the radius-only RandomForest in 99 of 100 splits (median $-0.026\text{ dex}$, $[-0.045,\,-0.008]$). Because these resampled partitions draw from the same 963 planets, they quantify sensitivity to the train-test split and confirm that our fiducial partition ($\Delta\mathrm{RMSE} = -0.038\text{ dex}$) is conservative relative to the median partition.

Because the fiducial split is made at the planet level, planets of the same system, which share their stellar parameters, can appear in both the training and the test set (Section 3.1). To test whether this affects the comparison, we repeated the 100 partitions with splits grouped by host star, so that all planets of a system fall in the same set, again stratified by mass class and with the same sample, models and hyperparameters. The advantage of RandomForest-multi over Baseline-PowerLaw is essentially unchanged (median $\Delta$RMSE $= -0.046$ dex, 2.5 to 97.5 percentile range $[-0.069, -0.016]$, lower RMSE in 99 of 100 splits, compared with $-0.048$ dex for planet-level splits), and the radius-only RandomForest has a lower RMSE than Baseline-PowerLaw in 94 of 100 splits (median $-0.022$ dex). The gain from the stellar and orbital covariates, measured by RandomForest-multi against the radius-only RandomForest, is smaller and less consistent (median $-0.022$ dex, negative in 94 of 100 splits). In the low-mass class its median falls from $-0.040$ dex for planet-level splits to $-0.026$ dex, it is negative in only 83 of 100 grouped splits, and its percentile range $[-0.120, +0.022]$ includes zero. The low-mass covariate gain reported in Section 5.1 is therefore not robust to grouping by host star. Table \ref{tab:robust} also lists one host-grouped split scored in the same way as the other variations.

\begin{table*}
\centering
\caption{Sensitivity analysis across sample selection, mass-boundary, feature, and train/test split perturbations using fixed hyperparameters. The host-grouped split has a different test set, so its RMSE values are not directly comparable with the fiducial row. Here $\Delta$ denotes $\mathrm{RMSE}(\text{best ML}) - \mathrm{RMSE}(\text{power law})$ with unadjusted 95\% paired bootstrap confidence intervals.}
\small
\setlength{\tabcolsep}{3pt}
\begin{tabular}{lclccl}
\hline
Variation & $N$ & Best full-coverage & RMSE & Power law & $\Delta$ (95\% CI) \\
\hline
fiducial (fixed hyperpar.) & 963 & RandomForest-multi & 0.327 & 0.368 & $-0.040$ $[-0.074, -0.007]$ \\
cuts: strict (0.20/0.06) & 759 & RandomForest-multi & 0.293 & 0.369 & $-0.076$ $[-0.110, -0.042]$ \\
cuts: loose (0.35/0.12) & 1161 & RandomForest-multi & 0.344 & 0.389 & $-0.045$ $[-0.079, -0.011]$ \\
boundaries: \cite{chen2016probabilistic}[13] & 963 & MLP & 0.315 & 0.341 & $-0.026$ $[-0.048, -0.003]$ \\
boundaries: wide low-mass & 963 & RandomForest-multi & 0.345 & 0.370 & $-0.025$ $[-0.065, +0.012]$ \\
sample: keep imaging & 974 & RandomForest-multi & 0.355 & 0.414 & $-0.059$ $[-0.099, -0.020]$ \\
features: radius only & 963 & MLP & 0.347 & 0.368 & $-0.020$ $[-0.043, +0.003]$ \\
features: no irradiation & 963 & RandomForest-multi & 0.333 & 0.368 & $-0.034$ $[-0.069, +0.003]$ \\
split: grouped by host star & 963 & RandomForest-multi & 0.301 & 0.345 &$ -0.045$ $[-0.070, -0.016]$\\
\hline
\end{tabular}
\label{tab:robust}
\end{table*}

\subsection{Feature attribution}
\label{sec:results:features}\label{sec:results:interp}
\begin{figure}[ht]
\centering
\includegraphics[width=0.5\linewidth]{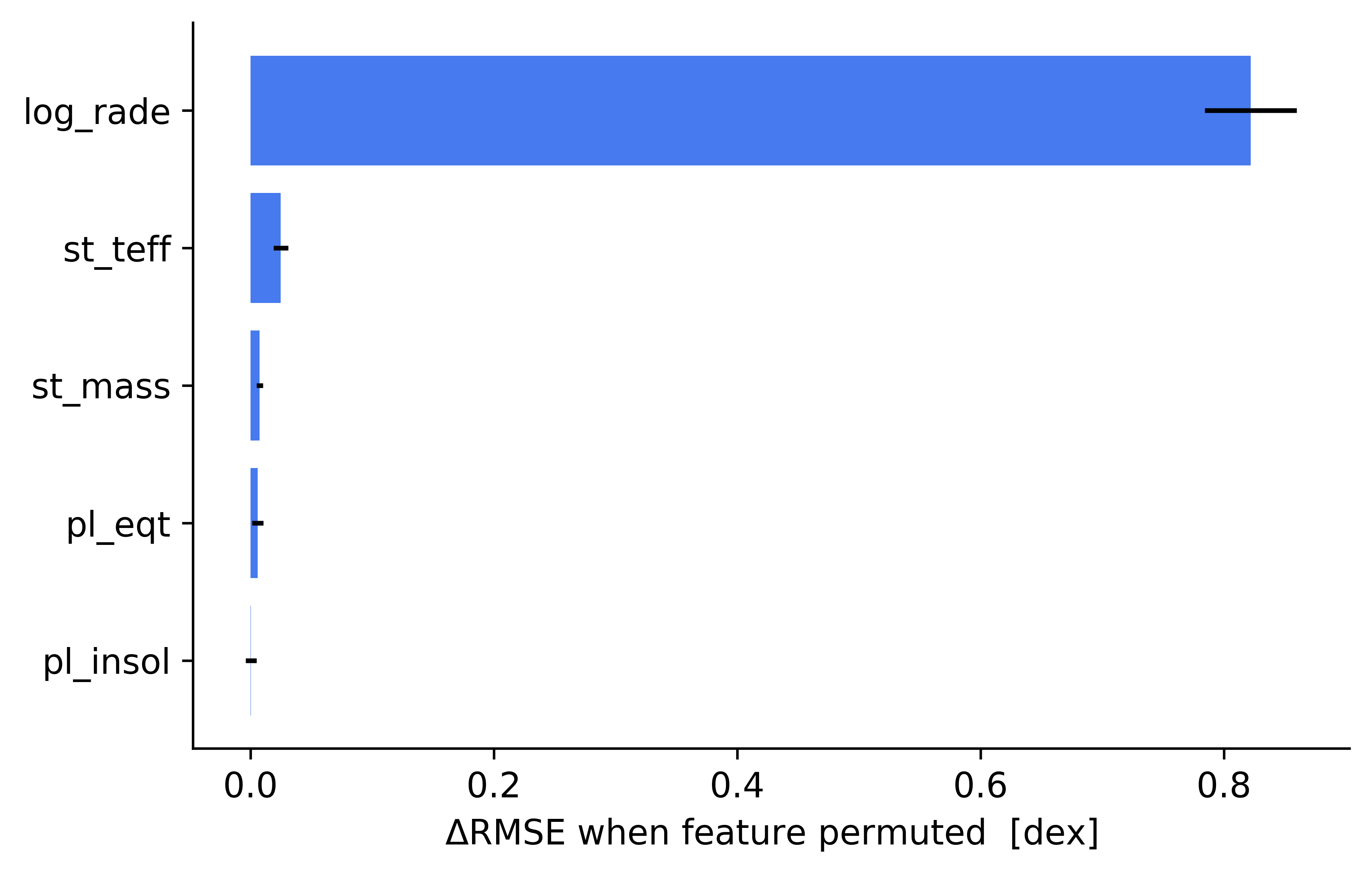}
\caption{Permutation feature importance for RandomForest-multi on the held-out test set, measured as the increase in test RMSE (in dex) when each feature is randomly shuffled.}
\label{fig:feat}
\end{figure}

Figure~\ref{fig:feat} shows the permutation feature importance for RandomForest-multi on the held-out test set. Planetary radius ($\log_{10} R$) dominates the predictions, with random permutation increasing test RMSE by $\Delta\mathrm{RMSE} = 0.822\text{ dex}$, whereas permuting stellar mass, stellar effective temperature, equilibrium temperature, or insolation flux changes the test RMSE by at most $0.025\text{ dex}$. Consistent with the paired covariate tests in Section~\ref{sec:results:overall}, secondary host-star and orbital parameters provide only marginal predictive leverage once planetary radius is specified.

\subsection{Selection effects and residual scatter decomposition}
\label{sec:results:bias}
\begin{figure}[ht]
\centering
\includegraphics[width=0.5\linewidth]{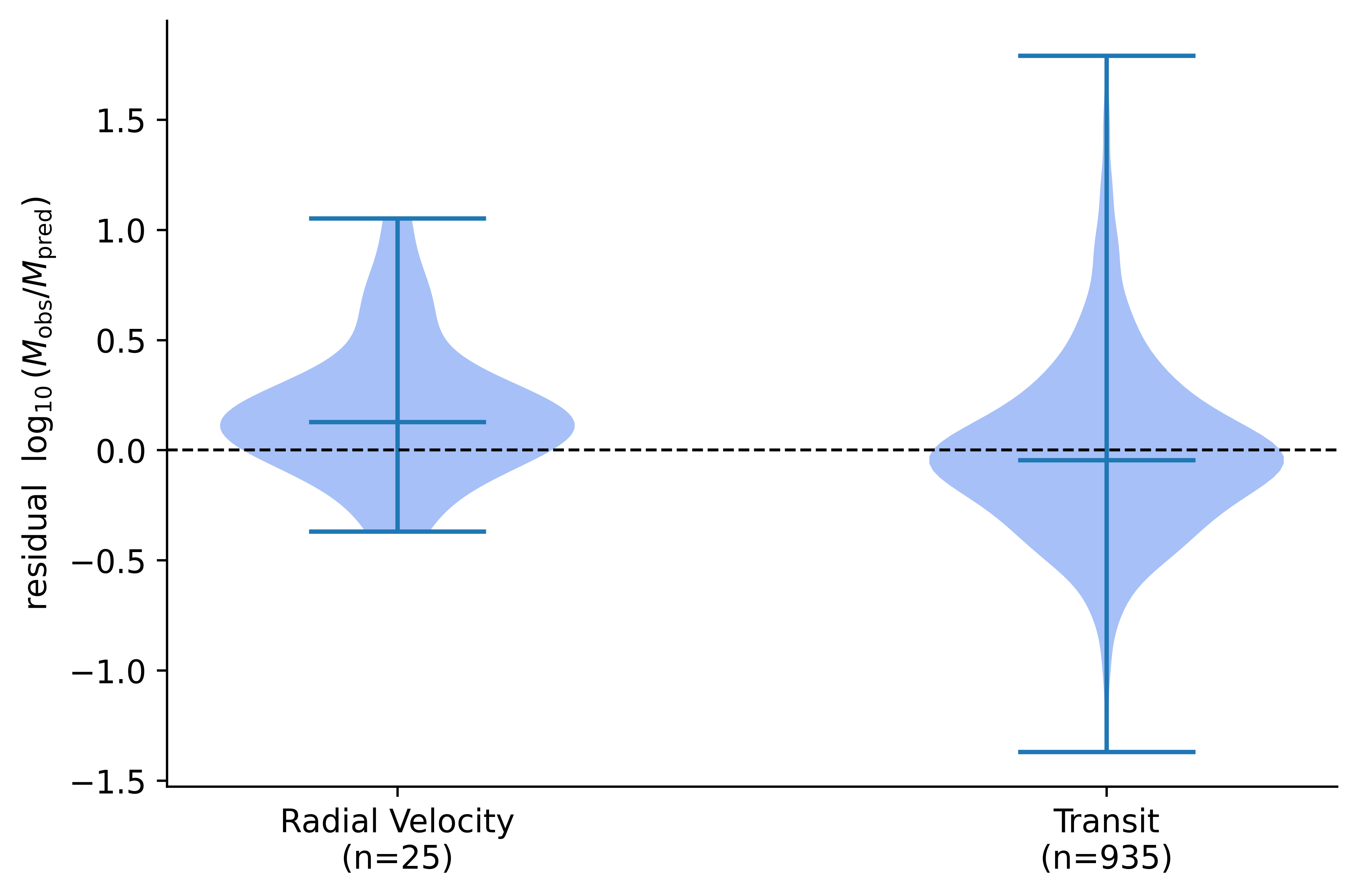}
\caption{Residual log-mass distributions grouped by discovery method. The RV-discovered and other transit-discovered subsamples differ in median residual by $0.173\text{ dex}$ ($95\%$ bootstrap CI $[0.112,\,0.261]$; Section~\ref{sec:results:bias}).}
\label{fig:bias}
\end{figure}

To examine observational selection effects, we compare $M$-$R$ residuals across discovery channels, requiring a minimum of 8 planets per group (Figure~\ref{fig:bias}). Because the NASA Exoplanet Archive table does not provide a uniform mass-measurement technique flag across all entries, discovery method serves here as an observational proxy; only three planets in the final sample were discovered through TTVs, which is insufficient to repeat the RV versus TTV offset analysis of \cite{mills2017planetary}[6]. Comparing the two well-populated discovery groups, namely RV-discovered planets ($N = 25$) and other transit-discovered planets, we find that their median mass residuals differ by $0.173\text{ dex}$ ($95\%$ bootstrap CI $[0.112,\,0.261]$, two-sample Kolmogorov-Smirnov $p = 0.0013$). At fixed radius, RV-discovered planets lie systematically above the mean relation, as expected from RV semi-amplitude detection thresholds that favor more massive planets. Selection effects also shape the orbital-period distribution of the benchmark: only 36 of the 963 planets have orbital periods $P > 100\text{ days}$, and only 4 of those have $M < 10\,M_\oplus$, reflecting the observational difficulty of securing precise dynamical masses for long-period, low-mass transiting planets.

Finally, we perform a first-order decomposition of the total residual scatter about an ordinary-least-squares power law fitted to all 963 benchmark planets ($\log_{10} M = 2.161 \log_{10} R + 0.086$, which yields a total residual standard deviation of $0.401\text{ dex}$). For each planet $i$, we propagate the catalog mass and radius uncertainties into the log-mass residual variance via

\begin{equation}
\sigma_i^2 = \left(\frac{\sigma_{M,i}}{M_i \ln 10}\right)^2 + a^2 \left(\frac{\sigma_{R,i}}{R_i \ln 10}\right)^2 ,
\end{equation}
where $a = 2.161$ is the power-law slope. Averaging $\sigma_i^2$ across all 963 planets gives a root-mean-square measurement uncertainty contribution of $0.067\text{ dex}$, accounting for only $\sim$3\% of the total residual variance ($0.067^2 / 0.401^2 \approx 0.028$). Subtracting the measurement variance in quadrature leaves $\sim$0.396~dex of non-measurement scatter. Because the reference model is a single power law, this remaining scatter combines intrinsic compositional diversity, survey selection biases, and departures of the true $M$-$R$ relation from log-linearity.

\vspace{0.5cm}
\section{Discussion}
\label{sec:discussion}
We evaluate empirical, probabilistic, and ML relations using a single homogeneous sample. The results show that much of the apparent disagreement in the $M$-$R$ literature arises from differences in catalog selection, validity domains, and evaluation metrics, while the choice of regression algorithm has a comparatively smaller effect. When evaluated on identical test planets across the full mass spectrum, four distinct ML architectures trained on radius alone achieve virtually identical accuracy ($0.344$ to $0.353\text{ dex}$) and show no statistically significant improvement over a two-parameter power law ($0.368\text{ dex}$). Meanwhile, within the small- and intermediate-planet radius domains for which they were calibrated, empirical broken power laws \citep{weiss2014mass,bashi2017two,otegi2020revisited}[11,14,9] and hierarchical probabilistic models \citep{chen2016probabilistic,parviainen2024spright}[13,23] match or outperform the ML regressors.

The results help define the applicability of ML methods to exoplanet mass estimation. ML models provide a single non-parametric predictor across the full radius range. This approach avoids the sharp boundary discontinuities associated with piecewise power laws. Permutation importance shows that radius accounts for almost all predictive power in the current archival sample. Furthermore, point ML models require external calibration, such as split conformal prediction, to provide reliable uncertainty intervals. Therefore, low-dimensional physical and probabilistic parameterizations remain preferable when predictions are restricted to low- and intermediate-radius planets.

In the giant-planet mass range, adding stellar and orbital parameters does not yield a statistically significant improvement over the radius-only model. This result may partly reflect the diversity of irradiation environments among giant planets. Strongly irradiated giant planets can exhibit substantial radius inflation, whereas this effect is generally weaker among less irradiated planets. Since the present analysis evaluates giant planets as a single mass-defined group, possible differences between these irradiation regimes may not be captured by the global performance metrics. Moreover, present-day equilibrium temperatures and insolation levels do not necessarily represent the cumulative irradiation received throughout a planet's evolutionary history. Planetary migration and changes in orbital configuration may further complicate the relationship between current orbital properties and planetary radius. Therefore, the limited predictive gain from the included stellar and orbital parameters does not necessarily indicate that irradiation is physically unimportant, but may reflect the limitations of using present-day covariates in a global prediction framework.

The variance decomposition provides additional constraints on future observational and modeling strategies. Because formal measurement errors contribute only $\sim$3\% of the residual variance in this quality-controlled sample ($\sigma_M/M < 0.25$, $\sigma_R/R < 0.08$), improving the precision of individual mass and radius measurements for already well-characterized planets will not substantially reduce the observed $\sim$0.396~dex scatter in the $M$-$R$ plane. Reducing or explaining this scatter requires more flexible functional models, including non-linear or multi-branch probabilistic relations that capture the giant-planet radius plateau. Hierarchical population frameworks are also needed to jointly account for intrinsic compositional dispersion and transit and RV survey completeness.
\\
\\
Several limitations should be considered when interpreting our benchmark results. First, because our sample requires both a transit radius and a precise dynamical mass ($\sigma_M/M < 0.25$), it inherits the combined selection functions of transit and RV follow-up surveys, including the scarcity of long-period ($P > 100\text{ days}$), low-mass planets and the systematic mass offset between RV and transit discovery channels quantified in Section \ref{sec:results:bias}. The resulting benchmark reflects the predictive performance for the subset of the exoplanet population accessible to current transit and RV follow-up surveys.

The residual offset between the groups of planets initially detected through radial-velocity observations and those initially detected through transit observations should also be interpreted with caution. Although the residuals are defined relative to a mass-radius relation and therefore account for the first-order dependence on planetary radius, the two groups could still have different orbital period and irradiation distributions. In particular, irradiation-driven radius inflation in gas giants may contribute to differences in the residual distributions at a given radius. The RV detection threshold also favors more massive planets, while the transit and RV discovery channels have different observational selection functions. Since the RV-discovered subsample contains only 25 planets, the present sample does not allow us to disentangle these effects or determine whether the observed offset persists after accounting for orbital period and irradiation. A larger and more uniformly characterized sample will be needed to investigate these contributions separately.

Second, a structural asymmetry exists between the evaluation of our newly trained models and that of published literature relations. Our ML regressors and baseline power laws are fitted strictly on the 723-planet training split and evaluated out of sample on the 240 held-out test planets. By contrast, we evaluate the relations of \cite{weiss2014mass}[11], \cite{chen2016probabilistic}[13], \cite{bashi2017two}[14], \cite{otegi2020revisited}[9], and \cite{parviainen2024spright}[23] using the coefficients reported in the original studies or their corresponding pre-trained software packages. Because those relations were calibrated on earlier subsets of the exoplanet archive that partially overlap with our 240 test planets, their test-set scores measure performance on a contemporary archival sample and may benefit mildly from this partial overlap. Our primary inferential comparisons between the ML regressors and the fitted power-law baseline are unaffected by this overlap, as both are trained and tested on the exact same data partitions. Accordingly, these two types of comparisons should be interpreted separately, as the newly trained models are evaluated on planets excluded from their training process, whereas the published relations are evaluated against the test sample.

Third, evaluating published empirical relations in a radius-only prediction task ($R \to M$) requires algebraic or domain adaptations when the original study parameterized radius as a function of mass \citep{bashi2017two}[14] or separated planetary branches using bulk density \citep{otegi2020revisited}[9]. Appendix \ref{app:empirical} provides the equations and boundary conditions adopted for these adaptations.

\vspace{0.5cm}
\section{Conclusions}
\label{sec:conclusions}
We have presented a homogeneous benchmark of exoplanet mass-radius relations using 963 planets with dynamical masses and transit radii retrieved from the NASA Exoplanet Archive. Our main conclusions are as follows:

A) Relative model accuracy depends strongly on planetary mass range, and no single relation is optimal across all masses. Over the full test set, the multi-feature random forest achieves the lowest RMSE ($0.329\text{ dex}$ versus $0.368\text{ dex}$ for a fitted single power law), but this difference ($-0.038\text{ dex}$, unadjusted $p = 0.007$, Holm-adjusted $p = 0.059$) is modest, and radius-only ML regressors ($0.344$ to $0.353\text{ dex}$) are statistically indistinguishable from the two-parameter power law. Scored in radius ranges, which can be used when the mass is unknown, the choice of method matters most below $1.6\,R_\oplus$, where the relation of \cite{weiss2014mass}[11], MLP, SVR, SPRIGHT and the relation of \cite{otegi2020revisited}[9] are significantly better than the power law, and least between 1.6 and $4\,R_\oplus$, where no method is.

B) Within the low- and intermediate-mass ranges ($M < 127\,M_\oplus$), published empirical relations \citep{weiss2014mass,bashi2017two,otegi2020revisited}[11,14,9] and SPRIGHT \citep{parviainen2024spright}[23] achieve lower or comparable errors ($0.142$ to $0.237\text{ dex}$) relative to ML models, and Forecaster \citep{chen2016probabilistic}[13] does so far low-mass planets ($0.142\text{ dex}$). On a matched domain of 96 small planets, SPRIGHT and Forecaster attain indistinguishable CRPS values ($0.113$ and $0.117\text{ dex}$) that both outperform a constant-variance probabilistic power law ($0.141\text{ dex}$).

C) Split conformal prediction provides ML point regressors with calibrated marginal coverage ($0.69$ at the 68\% level and $0.98$ at the 95\% level). Conditional coverage, however, varies across mass classes and radius ranges (over-covering low-mass planets and planets below $4\,R_\oplus$, and under-covering gas giants and planets above $4\,R_\oplus$), which motivating radius-stratified Mondrian calibration \citep{vovk2005algorithmic}[29] for future deployment.

D) Formal measurement uncertainties account for only $\sim$3\% of the residual variance ($0.067\text{ dex}$ of the $0.401\text{ dex}$ total scatter about a single power law), demonstrating that the remaining $\sim$0.396~dex of scatter is governed by intrinsic compositional degeneracy, survey selection biases, and functional approximation.

The benchmark established in this work provides a homogeneous framework for evaluating the predictive performance of mass-radius relations and assessing their applicability to exoplanet populations beyond the sample considered here.

\vspace{0.5cm}
\section*{Data Availability}
The benchmark catalog of 963 planets with all parameters used in this work, the identifiers of the 723 training and 240 test planets, the NASA Exoplanet Archive query result from which the catalog was built, and the complete analysis code are published in machine-readable form as a supplement to the online version of this study. The code rebuilds the sample, trains and evaluates every model, and regenerates the numbers, tables, and figures in this paper, including the radius-range and host-grouped analyses, with pinned software versions and fixed random seeds. The archive also contains the outputs of both the run reported in this study and a full run, including the predicted mass of every test planet for every method.

\vspace{0.5cm}
\section*{Acknowledgments}
The authors thank all individuals who contributed to this work through scientific discussions and constructive feedback. This research has made use of the NASA Exoplanet Archive, which is operated by the California Institute of Technology, under contract with the National Aeronautics and Space Administration under the Exoplanet Exploration Program. Astropy \citep{astropy2022}[30], NumPy, pandas, Matplotlib, SciPy, scikit-learn, Forecaster, MRExo, and SPRIGHT were used or evaluated in this work. AI-assisted tools were used to support code scaffolding, sentence polishing (improving sentence structure and readability), and language editing; all scientific content, analysis choices, and conclusions are the authors' own, and the authors take full responsibility for the work.

\vspace{0.7cm}
\appendix
\section{Hyperparameter grids and selected configurations}
\label{app:hyper}
Each machine-learning model in the benchmark is tuned via five-fold cross-validation restricted to the 723-planet training set, with the 240 held-out test planets excluded from all stages of model selection. Table~\ref{tab:hyper} lists the hyperparameter search grids and the optimal configuration selected for each model; all unlisted parameters remain at their default values in scikit-learn.

\begin{table*}
\centering
\caption{Hyperparameter search grids and configurations selected by five-fold cross-validation on the training set. The final column lists the inner cross-validation RMSE (CV RMSE, in dex of $\log_{10} M$) on the training split.}
\small
\setlength{\tabcolsep}{3pt}
\begin{tabular}{llll}
\hline
Model & Grid & Selected & CV RMSE \\
\hline
RandomForest &
\parbox[t]{0.30\textwidth}{\raggedright n\_estimators: 300; max\_depth: None, 6, 12; min\_samples\_leaf: 1, 3, 10} &
\parbox[t]{0.28\textwidth}{\raggedright max\_depth=6; min\_samples\_leaf=10; n\_estimators=300} &
0.392 \\

GradientBoosting &
\parbox[t]{0.30\textwidth}{\raggedright n\_estimators: 200, 400; learning\_rate: 0.03, 0.1; max\_depth: 2, 3} &
\parbox[t]{0.28\textwidth}{\raggedright max\_depth=2; n\_estimators=200; learning\_rate=0.03} &
0.386 \\

SVR &
\parbox[t]{0.30\textwidth}{\raggedright C: 1.0, 10.0, 100.0; gamma: scale, 0.1; epsilon: 0.05, 0.1} &
\parbox[t]{0.28\textwidth}{\raggedright C=100; epsilon=0.1; gamma=scale} &
0.381 \\

MLP &
\parbox[t]{0.30\textwidth}{\raggedright hidden\_layer\_sizes: (64,), (64, 64); alpha: 0.0001, 0.01; learning\_rate\_init: 0.001, 0.01} &
\parbox[t]{0.28\textwidth}{\raggedright alpha=0.0001; hidden\_layer\_sizes=(64, 64); learning\_rate\_init=0.01} &
0.377 \\

RandomForest-multi &
\parbox[t]{0.30\textwidth}{\raggedright n\_estimators: 300; max\_depth: None, 6, 12; min\_samples\_leaf: 1, 3, 10} &
\parbox[t]{0.28\textwidth}{\raggedright max\_depth=6; min\_samples\_leaf=1; n\_estimators=300} &
0.359 \\
\hline
\end{tabular}
\label{tab:hyper}
\end{table*}

\section{Empirical relations: functional forms, units, and validity domains}
\label{app:empirical}
Each empirical relation in the benchmark predicts planetary mass from radius using the equations and radius validity domains (in Earth units) summarized in Table~\ref{tab:empirical}. Outside its stated radius domain, a relation is recorded as inapplicable without extrapolation, resulting in the test-set coverage fractions reported in Table~\ref{tab:lookup}.

\begin{table*}
\centering
\caption{Empirical mass-radius relations and radius validity domains implemented in the benchmark.}
\footnotesize
\setlength{\tabcolsep}{4pt}
\begin{tabular}{lll}
\hline
Relation & Form ($M$ in $M_\oplus$, $R$ in $R_\oplus$) & Domain ($R_\oplus$) \\
\hline
\cite{weiss2014mass}[11] & $\rho=2.43+3.39R$ (g\,cm$^{-3}$) for $R<1.5$; $M=2.69R^{0.93}$ for $R\ge 1.5$ & $0.5$ to $4.0$ \\
\cite{otegi2020revisited}[9] (radius-only) & $M=\min\left(0.90R^{3.45},\,1.74R^{1.58}\right)$ & $0.5$ to $14.6$ \\
\cite{bashi2017two}[14] (inverted) & $M=124\,(R/12.1)^{1/0.55}$ & $0.5$ to $12.1$ \\
\hline
\end{tabular}
\label{tab:empirical}
\end{table*}

\noindent For \cite{weiss2014mass}[11], the published relation directly expresses bulk density as a function of radius for $R < 1.5\,R_\oplus$ (which we convert to mass via $M = \tfrac{4}{3}\pi R^3\rho$ in Earth units) and mass directly as a function of radius ($M = 2.69R^{0.93}$) for $1.5 \le R \le 4.0\,R_\oplus$. Because \cite{weiss2014mass}[11] specify an upper limit of $4.0\,R_\oplus$ without an explicit lower cutoff, we adopt $0.5\,R_\oplus$ as the lower domain boundary across all three empirical relations.

Whereas \cite{weiss2014mass}[11] formulated their relation directly for predicting mass from radius, the relations of \cite{otegi2020revisited}[9] and \cite{bashi2017two}[14] require mathematical adaptation for a radius-only benchmark ($R \to M$). Specifically, \cite{otegi2020revisited}[9] fitted two separate mass-as-a-function-of-radius power laws below $120\,M_\oplus$: $M = (0.90 \pm 0.06)R^{3.45 \pm 0.12}$ for rocky planets with bulk density $\rho > 3.3\text{ g cm}^{-3}$, and $M = (1.74 \pm 0.38)R^{1.58 \pm 0.10}$ for volatile-rich planets with $\rho < 3.3\text{ g cm}^{-3}$. Because bulk density depends on the unknown planetary mass and is unavailable when predicting mass from radius alone, our radius-only adaptation transitions between the two branches at the intersection radius where the two published curves cross, $R_{\times} = (1.74/0.90)^{1/(3.45-1.58)} = 1.42\,R_\oplus$, which is equivalent to taking the minimum of the two expressions in Table~\ref{tab:empirical}. Using a fixed radius threshold inevitably misclassifies some planets within the $2$ to $3\,R_\oplus$ compositional transition zone described by \cite{otegi2020revisited}[9]. To enforce their upper calibration limit of $120\,M_\oplus$ self-consistently from radius alone, we treat the relation as inapplicable wherever its predicted mass exceeds $120\,M_\oplus$, corresponding to $R > 14.6\,R_\oplus$ on the volatile-rich branch.

In contrast, \cite{bashi2017two}[14] parameterized planetary radius as a function of mass, reporting $R \propto M^{0.55 \pm 0.02}$ below a transition mass of $124 \pm 7\,M_\oplus$ (corresponding to a transition radius of $12.1 \pm 0.5\,R_\oplus$) and $R \propto M^{0.01 \pm 0.02}$ above it, without tabulating a separate intercept constant. Because the two power-law segments meet at the fitted breakpoint $(124\,M_\oplus,\,12.1\,R_\oplus)$, the normalization of the low-mass branch is uniquely fixed by that breakpoint. Algebraically solving their low-mass relation for mass as a function of radius yields $M = 124\,(R/12.1)^{1/0.55}$ (or $M \approx 1.33\,R^{1.82}$ in Earth units) over $0.5 \le R \le 12.1\,R_\oplus$. Above $12.1\,R_\oplus$, the high-mass exponent of $0.01 \pm 0.02$ reported by \cite{bashi2017two}[14] is statistically indistinguishable from zero, reflecting the physical insensitivity of giant-planet radii to mass; consequently, the upper branch cannot be inverted to predict mass from radius and is excluded outside $R \le 12.1\,R_\oplus$.

For the two probabilistic packages, SPRIGHT is queried through its public software interface, which returns posterior mass distributions over the radius interval $0.5 \le R \le 6.5\,R_\oplus$ and is recorded as inapplicable outside that range. Forecaster \citep{chen2016probabilistic}[13] is executed using the authors' public implementation across its full calibrated mass domain of $3 \times 10^{-4}$ to $3 \times 10^{5}\,M_\oplus$, providing predictions for all 240 test planets. For both packages, the point mass used in the RMSE is the median of the predictive draws in $\log_{10}M$, which equals the logarithm of the median in linear mass. Finally, the non-parametric MRExo package \citep{kanodia2023beyond}[17] is not included in the numerical tables because its pre-trained lookup tables are calibrated specifically for planets orbiting M-dwarf hosts rather than the heterogeneous sample of FGKM host stars analysed in this benchmark.


\end{document}